\documentclass[aps, prb, showpacs, preprint, preprintnumbers, groupedaddress]{revtex4-1}

\usepackage{graphicx}
\usepackage{rotating}
\usepackage{dcolumn}

\begin{document}


\title{Anisotropic thermal expansion of bismuth from first principles}

\author{B.~Arnaud$^{1}$, S. Leb\`egue$^{2}$ and G. Raffy$^{1}$}

\affiliation{$^{1}$Institut de Physique de Rennes (IPR),
UMR UR1-CNRS 6251, Campus de Beaulieu - Bat 11 A, 35042 Rennes Cedex, France, EU}

\affiliation{$^{2}$Laboratoire de Cristallographie, R\'esonance Magn\'etique et Mod\'elisations (CRM2), 
UMR CNRS 7036, Institut Jean Barriol, Universit\'e de Lorraine, BP 239, Boulevard des Aiguillettes, 54506 Vandoeuvre-l\`es-Nancy, France, EU}




\date{\today}

\begin{abstract}
Some anisotropy in both mechanical and thermodynamical properties of bismuth is expected.
A combination of density functional theory
total energy calculations and density functional perturbation theory in the local density approximation 
is used to compute the elastic constants at 0 K using a finite strain approach 
and the thermal expansion tensor in the quasiharmonic approximation.  The overall agreement with experiment is good. 
Furthermore, the anisotropy in the thermal expansion is found to arise from the anisotropy in both the directional
compressibilities and the directional Gr\"uneisen functions.

\end{abstract}


\maketitle

\section{Introduction\label{intro}}
The semimetal bismuth is of interest both scientifically and technologically. Indeed, it exhibits many fascinating properties,
like giant magnetoresistance\cite{alers_1953, xu_2005}, thermoelectricity\cite{Boydston_1927, chandrasekhar_1959}, 
large diamagnetism\cite{fuseya_2015} that can be ascribed 
to the peculiar electronic structure of bismuth, namely the small overlap between the valence and the conduction bands giving rise 
to a Fermi surface made of tiny electron and hole pockets\cite{hofmann_2006}. Real-world applications of bismuth related to the aforementioned properties
range from hall magnetometry\cite{novoselov_2003} to diamagnetic levitation on the microscale\cite{kokorian_2014}.

The possibility to drive bismuth strongly out of equilibrium by an ultrashort laser pulse is also behind a huge
amount of experimental\cite{sokolowski_2003, fritz_2007, johnson_2008, johnson_2009, papalazarou_2012, faure_2013} 
and theoretical work\cite{murray_2005, zijlstra_2006, giret_2011, arnaud_2013, murray_2015}. 
From a theoretical point of view, the ultrafast dynamics of coherent optical phonons has been tackled by means 
of first-principles calculations where the lattice parameters are kept constant. However, the development of strain from coherent
acoustic phonons on a picosecond timescale is still poorly understood\cite{laulhe_2013} and has never been addressed by ab-initio calculations.
A prerequisite for achieving such a goal is to demonstrate that the thermal expansion of bismuth at equilibrium can be understood
and predicted by performing ab-initio calculations. A good strategy is to resort to the quasiharmonic approximation,
where the atoms of the crystal are considered to undergo harmonic oscillations, but with frequencies that depend on strain. This
approximation, when combined with density functional perturbation theory\cite{giannozzi_1991}, has been found to produce thermal expansion coefficients
in good agreement with experimental results well below the melting temperature of isotropic\cite{debernardi_1996, karki_1999, debernardi_2001} 
and anisotropic solids\cite{lichtenstein_1998, schelling_2003, mounet_2005, wang_2006}. 

The paper is organized as follows. In section \ref{details}, we give an account of the technicalities used
to perform our first-principles calculations. 
In section \ref{structure}, we describe the crystallographic structure
of bismuth and compare our calculated lattice constants with and without spin-orbit interaction (SOI) 
to the experimental lattice constants at 4 K obtained from X-ray measurements.
In section \ref{elastic_properties}, we explain how the elastic constants at 0 K can be computed
using a finite strain method and make a comparison with available experimental results indirectly obtained by measuring the
sound wave velocities for different directions and polarizations. We also discuss the impact of SOI on the calculated elastic 
constants. In section \ref{expansion}, we introduce the theory 
allowing the calculation of the thermal expansion tensor of bismuth within the quasiharmonic approximation and compare the
thermal expansion coefficients parallel and perpendicular to the trigonal axis to some measurements made using an optical lever
dilatometer. We also unravel the role respectively played by the elasticity and the anharmonicity in the anisotropy of the
thermal expansion coefficients of bismuth. Finally, the specific heat at constant pressure is reported and compared
with calorimetry measurements.

\section{Computational details}\label{details}
All the calculations are performed using the ABINIT code\cite{gonze_2009}. We use
a plane-wave basis set, the Hartwigsen-Goedecker-Hutter (HGH) pseudopotentials\cite{goededecker_1996}
and the local density approximation (LDA) for the exchange-correlation functional. We carefully
check the convergence of our results with respect to the wave function cut-off and the $k$-point sampling
of the Brillouin zone. A 40 Ry cut-off and a $16\times16\times16$ mesh for Brillouin zone sampling ensure
that our results (lattice parameters, elastic constants, phonon frequencies) are well converged. The
dynamical matrix is explicitly calculated on a $8\times8\times8$ $q$-point mesh 
using density functional perturbation theory\cite{gonze_1997} and the phonon frequencies
are Fourier interpolated on a $32\times 32\times 32$ $q$-point mesh in order to compute
the thermal expansion tensor. We include SOI in all calculations but also present the lattice parameters and
elastic constants obtained without SOI in order to highlight the crucial role played by SOI.

\section{Lattice parameters at zero temperature}\label{structure} 
\begin{table}
\caption{Calculated LDA lattice parameters with and without SOI compared to the experimental results 
of Ref\cite{cucka_1962}.} 
\label{lattice_tab}
\begin{tabular}{lccccc}
\hline
   &\multicolumn{3}{c}{Rhombohedral structure} &\multicolumn{2}{c}{Hexagonal structure} \\
\hline
 & $a_0$ (\AA) & $\alpha_0$ ($^o$) & $V_0$ (\AA$^3$) & $a_{\parallel,0}$ (\AA) & $a_{\perp,0}$ (\AA) \\ 
Experiment & 4.724 & 57.35 &  69.97   &11.796  & 4.533  \\
Theory (without SOI)  & 4.653    & 57.48 & 67.12     &11.610  & 4.475  \\     
Theory (with SOI)  & 4.697    & 57.53 & 69.10     &11.714  & 4.521  \\     
\hline
\end{tabular}
\end{table}
Bismuth crystallizes in a rhomboedral structure, also called A7 structure, with two atoms per
unit cell. The vectors spanning the unit cell are given by
\begin{equation}
{\bf{a}}_1=\left(a \xi, -\frac{a \xi}{\sqrt{3}}, h \right) ;
{\bf{a}}_2=\left(0, \frac{2 a \xi}{\sqrt{3}}, h \right) ;
{\bf{a}}_3=\left(-a \xi, -\frac{a \xi}{\sqrt{3}}, h \right),
\end{equation}
where $\xi=\sin[\frac{\alpha}{2}]$ and $h=a\sqrt{1-\frac{4}{3}\xi^2}$. The length
of the three lattice vectors is equal to $a$ and the angle between any pair of vector 
is $\alpha$. The two atoms belonging to the unit cell are located at 
$\pm u \left({\bf{a}}_1+{\bf{a}}_2+{\bf{a}}_3\right)$ where $u$ is a dimensionless parameter
and ${\bf{a}}_1+{\bf{a}}_2+{\bf{a}}_3$ is parallel to the ternary axis (C$_3$ axis). Alternatively,
the structure can be viewed as an hexagonal structure spanned by the following three lattice 
vectors
\begin{equation}
\tilde{\bf{a}}_1= {\bf{a}}_1 - {\bf{a}}_2 ;
~\tilde{\bf{a}}_2= {\bf{a}}_2 - {\bf{a}}_3;
~\tilde{\bf{a}}_3= {\bf{a}}_1+{\bf{a}}_2+{\bf{a}}_3,
\end{equation} 
where $\tilde{a}_1=\tilde{a}_2 \equiv a_\perp$ and $\tilde{a}_3 \equiv a_\parallel$. The lattice
cell parameters of the two structures are related to each other by the following relations
\begin{equation}
\begin{array}{lcl}
a_\perp=2 a \sin\left(\frac{\alpha}{2}\right) & & a=\frac{1}{3} \sqrt{3 a_\perp^2+ a_\parallel^2}              \\
a_\parallel=a\sqrt{3+6\cos(\alpha)} &  &  \sin\left[\frac{\alpha}{2}\right]=\frac{3}{2} a_\perp/\sqrt{3 a_\perp^2+ a_\parallel^2}
\end{array}
\end{equation}
All the calculations have been performed using the rhombohedral structure because it contains three times atoms less than the 
hexagonal structure. However, the hexagonal structure, as will be seen later, is more convenient to define thermal expansion
coefficients. 
Our calculated LDA lattice parameters with and without SOI are given in Table \ref{lattice_tab} along with the experimental results
at 4.2 K\cite{cucka_1962}. The agreement between theory and experiment is significantly improved when SOI 
is included\cite{diaz_2007}. Indeed, $a_{\parallel,0}$ and $a_{\perp,0}$ are
respectively underestimated from 0.69 \% (1.58 \%) and 0.26 \% (1.27 \%) with respect to experiments, 
leading to an underestimation of the equilibrium volume $V_0$ of 1.2 \% (4.1 \%) when SOI is included (neglected). Thus,
the inclusion of SOI is mandatory to achieve a better description of the equilibrium lattice parameters of bismuth. 
 
\section{Elastic constants at zero temperature}\label{elastic_properties}
\begin{figure}[h!]
\vskip1.0cm
\includegraphics[width=8.5cm]{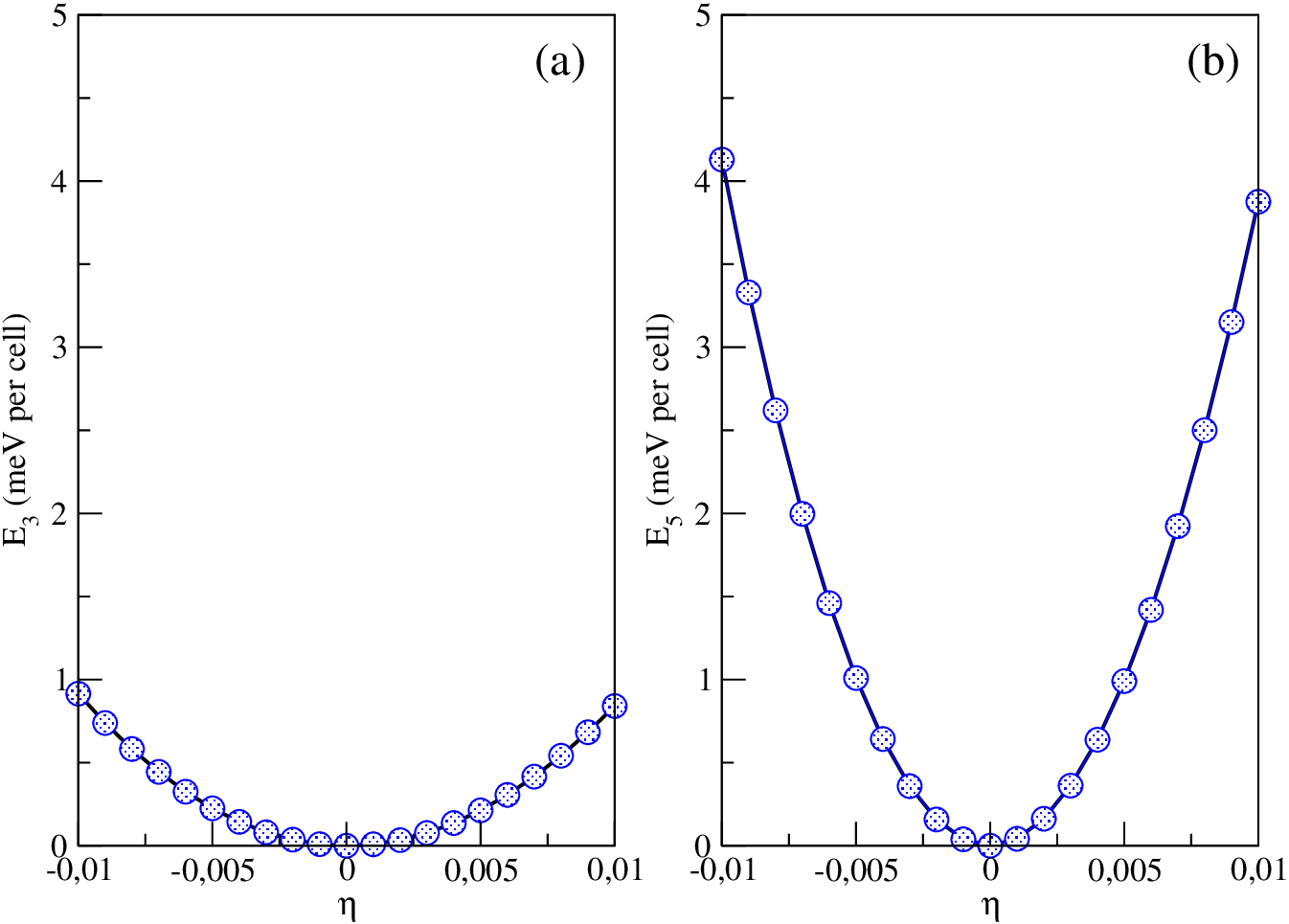}
\caption{\label{energy_vs_strain}
Energy per unit cell (in meV) as a function of $\eta$ (dimensionless quantity) for the deformations respectively labelled ${\bf {\eta}}_3$ 
 (Left panel) and ${\bf {\eta}}_5$ (Right panel). The circles denote the results of the LDA calculations including
SOI whereas the full lines are the result
of the polynomial fit according to Eq. \ref{fit_order4}. The zero of energy corresponds to the energy of the fully relaxed structure 
(unstrained reference structure of volume $V_0$).}
\end{figure}

The elastic properties of a bismuth crystal can be inferred from the theory of
elasticity. The Lagrangian strain tensor $\bf{\eta}$ is defined as
\begin{equation}\label{eta_def}
\eta_{ab}=\epsilon_{ab}+\frac{1}{2} \sum_k \epsilon_{ak}\epsilon_{kb}
\end{equation}
where $\bf{\epsilon}$ is the linear strain tensor which transforms a vector
$\bf{a}$ into $(\bf{1+\epsilon})\bf{a}$. The energy of the crystal per unit cell
$E$ can be expanded in power series with respect to the strain  $\bf{\eta}$ as
\begin{equation}\label{energy_1}
E[{\bf{\eta}}]=E_0 + \frac{V_0}{2} \sum_{ijkl} C_{ijkl} \eta_{ij} \eta_{kl} + \cdots
\end{equation}
where $E_0$ and $V_0$ are the energy and the volume of the unstrained unit cell
and $C_{ijkl}$ are the elastic stiffness constants of the crystal. Using Voigt's notation, Eq. \ref{energy_1} can be written as
\begin{equation}\label{energy_2}
E[{\bf{\eta}}]=E_0 + \frac{V_0}{2} \sum_{\alpha \beta} C_{\alpha \beta} \eta_{\alpha} \eta_{\beta} + \cdots
\end{equation}
where the fourth-rank stiffness tensor has been replaced by a $6\times 6$ matrix $\bf{C}$. 
By virtue of the rhombohedral structure A7 of bismuth (space group R$\overline{3}$m), the matrix  $\bf{C}$
can be cast in the form
\begin{equation}\label{elasticity_tensor}
{\bf{C}}= \left( \begin{array}{cccccc}      
C_{11} & C_{12} & C_{13} & C_{14} & 0 & 0 \\ 
C_{12} & C_{11} & C_{13} & -C_{14} & 0 & 0 \\
C_{13} & C_{13} & C_{33} & 0 & 0 & 0 \\
C_{14} & -C_{14} & 0 & C_{44} & 0 & 0 \\
0 & 0 & 0 & 0 & C_{44} & C_{14} \\
0 & 0 & 0 & 0 & C_{14} & C_{66} \\
\end{array} \right) 
\end{equation}
provided that the $z$ axis is taken along the trigonal axis. Here, 
only 6 elements are independent and $C_{66}=\frac{1}{2}\left[C_{11}-C_{12}\right]$. In order
to compute these elements, we consider six sets of deformations parametrized by $\eta$
\begin{equation}\label{distorsions}
\begin{array}{lll}
{\bf {\eta}}_1 = ( \eta, \eta, \eta, 0 ,0, 0) &; {\bf {\eta}}_2 = ( \eta, 0, 0, 0 ,0, 0) &; {\bf {\eta}}_3 =  ( 0, 0, \eta, 0 ,0, 0) \\
{\bf {\eta}}_4 = ( 0, 0, 0, 2\eta ,0, 0) &; {\bf {\eta}}_5 = ( \eta, \eta, 0, 0 ,0, 0)&; {\bf {\eta}}_6  =  ( \eta, 0, 0, 2\eta ,0, 0),
\end{array}
\end{equation}
where $\eta$ is varied between -0.01 and 0.01 with a step of 0.001. For each deformation labelled by $i$ and 
each value of $\eta$, we
build the Lagrangian strain matrix ${\bf{\eta}}$ and solve Eq. \ref{eta_def} in an iterative way to obtain 
the matrix ${\bf{\epsilon}}$. 
Thus, we generate a distorted cell from the undistorted one by using the matrix ${\bf{\epsilon}}$
and we compute the total energy $E_i(\eta)\equiv E[{\bf{\eta}}_i]$ of the distorted structure
where the atomic positions are fully relaxed. Then, the energy per unit volume is fitted by
a polynomial of order 4
\begin{equation}\label{fit_order4}
\frac{E_i(\eta)}{V_0}=\sum_{j=0}^4 A_j^i \eta^j,
\end{equation}
where $A_0^i=E_0/V_0$, $A_1^i=0$ and $A_2^i$ can be expressed as a function of the unknown second order
elastic constants as
\begin{equation}\label{coeff2}
\begin{array}{lll}
A_2^1 = C_{11}+C_{12}+2 C_{13} +\frac{1}{2} C_{33} &; A_2^2  = \frac{1}{2} C_{11} &; A_2^3  =  \frac{1}{2} C_{33}\\
A_2^4 = 2 C_{44} &; A_2^5 = C_{11}+C_{12}   &; A_2^6 = \frac{1}{2} C_{11}+ 2 C_{14} + 2 C_{44}  \\
\end{array}
\end{equation}
by using Eq. \ref{energy_2} and \ref{elasticity_tensor}. The computed total energy per unit cell at 0 K 
including SOI (circles) together with the polynomial fit (full lines) are displayed in Fig. \ref{energy_vs_strain} for 
the two deformations respectively labelled ${\bf {\eta}}_3$ ($a_\parallel$ varies while keeping 
$a_\perp$ constant) and ${\bf {\eta}}_5$ ($a_\perp$ varies while keeping $a_\parallel$ constant).
The elastic constant $C_{33}/2$ (see Fig. \ref{energy_vs_strain}.a) is roughly four times smaller than $C_{11}+C_{12}$
(see Fig. \ref{energy_vs_strain}.b), reflecting the fact that bismuth is much softer against a strain along the trigonal
axis than against a strain perpendicular to it. We do not show the energy per unit cell for the four remaining 
deformations but compare in Table \ref{elasticity_tab} our calculated elastic constants with and without SOI to the experimental 
elastic constants indirectly obtained from the measurements of ultrasonic wave velocities by the 
pulse echo technique\cite{lichnowski_1976}.
\begin{table}
\caption{
Calculated LDA elastic constants with and without SOI of Bi compared to the experimental results
at 4.2 K of Ref. \cite{lichnowski_1976}.}
\label{elasticity_tab}
\begin{tabular}{ccccccc}
\hline
    & $C_{11}$ (GPa) & $C_{12}$  (GPa)&  $C_{13}$   (GPa)& $C_{14}$ (GPa) & $C_{33}$ (GPa)& $C_{44}$  (GPa)\\ 
\hline
Experiment\cite{lichnowski_1976}    & 69.3 & 24.5      & 25.4        &  8.40    & 40.4     & 13.5       \\
Present work (without SOI) & 84.6    &  30.2     &  27.9           &   9.8        &  46.1       &    16.0         \\
Present work  (with SOI) & 67.7    &  25.0     &  24.3           &   5.9        &  40.6       &    8.7         \\
\hline
\end{tabular}
\end{table}
The calculated elastic constants without SOI are all overestimated with respect to experiments. Taking into account
SOI leads to a decrease of all elastic constants that can be explained in a very qualitative way as follows:
SOI mixes bonding and antibonding states and not only increases the equilibrium volume
by 2.9 \% (see Table \ref{lattice_tab}) but also softens the elastic constants by as much as 45 \% (the larger effect being for
$C_{44}$). As shown in Table \ref{elasticity_tab}, the overall agreement between theory and experiment is significantly improved 
when SOI is included. The only exceptions are the
$C_{14}$ and $C_{44}$ elastic constants which are underestimated by about $30\%$ with respect to experiment. Such a discrepancy
might be ascribed to experimental uncertainties associated with the sample and its bonded transducer and/or to long
range effects, like van der Waals interactions, not captured by the LDA exchange-correlation functional\cite{bucko_2013}. 
It is also worth mentioning that all the calculated elastic constants with or without SOI satisfy Born's mechanical 
stability for a rhomboehedral structure\cite{born_1940, mouhat_2014} ensuring that bismuth is stable at 0 K. 

Inverting the matrix $\bf{C}$ defined in Eq. \ref{elasticity_tensor} leads to the following expression for
the compliance matrix :
\begin{equation}\label{compliance_tensor}
{\bf{S}}= \left( \begin{array}{cccccc}      
S_{11} & S_{12} & S_{13} & S_{14} & 0 & 0 \\ 
S_{12} & S_{11} & S_{13} & -S_{14} & 0 & 0 \\
S_{13} & S_{13} & S_{33} & 0 & 0 & 0 \\
S_{14} & -S_{14} & 0 & S_{44} & 0 & 0 \\
0 & 0 & 0 & 0 & S_{44} & 2 S_{14} \\
0 & 0 & 0 & 0 & 2 S_{14} & S_{66} \\
\end{array} \right) 
\end{equation}
where $S_{66}=2(S_{11}-S_{12})$.
The anisotropy in the elastic properties of Bi can be ascertained by introducing a parallel and a perpendicular compressibility 
respectively defined as
\begin{equation}\label{comp_par}
\chi_\parallel=-\frac{\partial \epsilon_{33}}{\partial P}=2S_{13}+S_{33}=(-2C_{13}+C_{11}+C_{12})/M
\end{equation}
and
\begin{equation}\label{comp_perp}
\chi_\perp=-\frac{\partial \epsilon_{11}}{\partial P}=-\frac{\partial \epsilon_{22}}{\partial P}=S_{11}+S_{12}+S_{13}=
(C_{33}-C_{13})/M,
\end{equation}
where $M=C_{33}(C_{11}+C_{12})-2C_{13}^2$.
\begin{table}
\caption{
Hydrostatic compressibilities $\chi$,  $\chi_\parallel$ and $\chi_\perp$ of Bi (in MBar$^{-1}$) obtained
by using the elastic constants computed with and without SOI (see Table \ref{elasticity_tab}) compared to the
values inferred from the experimental results at 4.2 K of Ref.\cite{lichnowski_1976}. 
The ratio $\chi_\parallel/\chi_\perp$ is also given to quantify the anisotropy of the elastic properties of Bi.}
\label{compressibility_tab}
\begin{tabular}{ccccc}
\hline
    & $\chi$  & $\chi_\parallel$ &  $\chi_\perp$   & $\chi_\parallel/\chi_\perp$ \\ 
\hline
Experiment\cite{lichnowski_1976}    & 2.92 & 1.72      &  0.60       &  2.86       \\
Present work  (without SOI) &  2.55   &  1.58     &  0.49    &   3.24       \\
Present work  (with SOI) & 2.97    &  1.71     &  0.63    &   2.71       \\
\hline
\end{tabular}
\end{table}
The computed values of $\chi_\parallel$ and $\chi_\perp$ with and without SOI are reported in Table \ref{compressibility_tab}.
The compressibilities are underestimated with respect to experiments when SOI is neglected. Taking into account SOI leads
to a very good agreement between the theoretical and experimental compressibilities in accordance with the fact that the 
elastic constants that play a role in the compressibilities are much better described 
when SOI is included (see Table \ref{elasticity_tab}).
The ratio of linear compressibilities $\chi_\parallel/\chi_\perp$ calculated when SOI is included indicates that the
contraction along the trigonal axis is about 2.7 times larger than the contraction perpendicular to it upon applying
an hydrostatic pressure. As stated before and illustrated in Fig. \ref{energy_vs_strain}, Bi is stiffer perpendicularly 
to the trigonal axis than parallel to it. The bulk modulus $B$ that measures material's resistance to uniform compression 
is defined as $B= -V_0\frac{\partial P}{\partial V} = 1/\chi$ where the hydrostatic compressibility $\chi$ is defined as
$\chi=\chi_\parallel+2\chi_\perp$.
Using the values of $\chi_\parallel$ and $\chi_\perp$ reported in Table \ref{compressibility_tab}, we obtain a theoretical
value of 33.65 (39.15) GPa for the bulk modulus $B$ when the SOI is included (neglected). Thus, the theoretical value including 
the SOI agrees well with the experimental value of 34.23 GPa at 4 K\cite{lichnowski_1976}. All the forthcoming calculations include SOI.

\section{Thermal expansion}\label{expansion}
We now present an analysis of the thermal expansion of bismuth using
Gr\"uneisen's theory. We follow the approach of Schelling and Keblinski\cite{schelling_2003}
and emphasize the differences in the formalism used to treat the thermal expansion in anisotropic and cubic solids. Since bismuth is an anisotropic solid, the
thermal expansion can be described in terms of the $3\times 3$ thermal
expansion tensor $\alpha$ whose components are written in terms of
the strain tensor $\epsilon$ as
\begin{equation}
\alpha_{ab}=\left( \frac{\partial \epsilon_{ab}}{\partial T}\right)_\sigma
\end{equation}
where $T$ is the temperature and where the subscript $\sigma$ means that
the temperature derivative is taken at constant stress. 
According to the generalized form of Hookes' law, we have:
\begin{equation}\label{compliance_eq}
\epsilon_{ab}=\sum_{de} S_{abde} \sigma_{de},
\end{equation}
where $\sigma_{de}$ and $S_{abde}$ are the stress and the compliance elastic tensor
respectively. Eq. \ref{compliance_eq} can be inverted and leads to
\begin{equation}\label{stiffness_eq}
\sigma_{de}=\sum_{ij} C_{deij} \epsilon_{ij},
\end{equation} 
where $C_{deij}$ is the stiffness elastic tensor. Thus, we have
\begin{equation}\label{inverse_eq}
\sum_{de} S_{abde} C_{deij} = \delta_{ai} \delta_{bj}
\end{equation}
By differentiating Eq. \ref{inverse_eq} with respect to $T$ and by using
Eq. \ref{compliance_eq} and \ref{stiffness_eq}, 
it is straightforward to show that the components of
$\alpha$ can be written as
\begin{equation}\label{def_alpha1}
\alpha_{ab}=-\sum_{de} \left( \frac{\partial \epsilon_{ab}}{\partial \sigma_{de}}\right)_T \times \left( \frac{\partial \sigma_{de}}{\partial T}\right)_\epsilon,
\end{equation}
where the temperature derivative of the stress tensor is taken at constant strain. 
This equation generalizes the expression
\begin{equation}
\alpha=\frac{1}{3B} \left(\frac{\partial P}{\partial T}\right)_V
\end{equation}
defining the thermal expansion coefficient $\alpha$ of a cubic solid
as a function of the bulk modulus $B$ and the partial derivative of the 
pressure with respect to temperature at constant volume. The stress tensor $\sigma_{de}$ appearing in Eq. \ref{def_alpha1} is defined as
\begin{equation}\label{def_sigma}
\sigma_{de}=\frac{1}{V_0} \frac{\partial F }{\partial \epsilon_{de}}
\end{equation}
where $F$ is the Helmholtz free energy per unit cell of the crystal defined as
\begin{equation}\label{def_fenergy}
F[{\bf{\epsilon}}]=E[{\bf{\epsilon}}] + F_{vib}[{\bf{\epsilon}}, T]=
E[{\bf{\epsilon}}]+\frac{1}{N} \sum_{{\bf{q}}, \lambda} \frac{ \hbar \omega_\lambda({\bf{q}})}{2}
+ k_B T \frac{1}{N} \sum_{{\bf{q}}, \lambda} \ln\left[1-\exp\left(-\frac{\hbar \omega_\lambda({\bf{q}})}{k_B T}\right)\right]
\end{equation}
where the electron entropy is discarded and the vibrational contribution $F_{vib}[{\bf{\epsilon}}, T]$ is computed
within the harmonic approximation. Here, $\omega_\lambda({\bf{q}})$ is the frequency of the phonon mode
$({\bf{q}}, \lambda)$ corresponding to wavevector ${\bf{q}}$ and polarization $\lambda$ and $N$ is the number of
qpoints included in the summation. Hence, combining Eq. \ref{def_sigma} and Eq. \ref{def_fenergy} leads to
\begin{equation}\label{def_sigma1}
\sigma_{de}=\frac{1}{V_0} \left[ \frac{\partial E }{\partial \epsilon_{de}} -\frac{1}{N} \sum_{{\bf{q}}, \lambda}
\gamma_{{\bf{q}}, \lambda}^{de} \hbar \omega_\lambda({\bf{q}})\left(n_{{\bf{q}}, \lambda}+\frac{1}{2}\right) \right],
\end{equation}
where $n_{{\bf{q}}, \lambda}$ is the Bose occupation factor at temperature $T$ for a phonon with frequency 
$\omega_\lambda({\bf{q}})$ and $\gamma_{{\bf{q}}, \lambda}^{de}$ is a generalized mode Gr\"uneisen parameter given by
\begin{equation}
\gamma_{{\bf{q}}, \lambda}^{de}=-\frac{\partial \ln \omega_\lambda({\bf{q}}) }{\partial \epsilon_{de}}
\end{equation} 
Note that $\sigma_{de}$ is temperature dependent because of the second term in Eq. \ref{def_sigma1} and
that $\sigma_{de}(T\to 0)$ is renormalized by zero point atomic motions. By derivating Eq. \ref{def_sigma1} with
respect to temperature $T$ at constant strain ${\bf{\epsilon}}$, we get
\begin{equation}\label{def_partial_sigma}
\left(\frac{\partial \sigma_{de}}{\partial T}\right)_\epsilon=- \sum_{{\bf{q}}, \lambda} \gamma_{{\bf{q}}, \lambda}^{de}
C_{{\bf{q}}, \lambda}
\end{equation}
where
\begin{equation}\label{def_specific_heat}
C_{{\bf{q}}, \lambda}=\frac{1}{N} \frac{\hbar \omega_\lambda({\bf{q}})}{V_0} \left(\frac{\partial n_{{\bf{q}}, \lambda} }{ \partial T  }\right)_\epsilon
\end{equation}
is the contribution of mode $({\bf{q}}, \lambda)$ to the lattice specific heat per unit volume at constant volume $C_V$. Thus $C_V$ is given by
\begin{equation}\label{def_specific_heat_2}
C_V(T)=\sum_{{\bf{q}}, \lambda}  C_{{\bf{q}}, \lambda} = \frac{k_B}{V_0} \frac{1}{N}\sum_{{\bf{q}}, \lambda}
\left(\frac{ \hbar \omega_\lambda({\bf{q}})}{2 k_B T} \right)^2 \frac{1}{\sinh^2\left(\frac{\hbar \omega_\lambda({\bf{q}})}{2 k_B T} \right)}.
\end{equation}
Finally, by using Eq. \ref{compliance_eq} and inserting Eq. \ref{def_partial_sigma} in Eq. \ref{def_alpha1}, we obtain
\begin{equation}\label{def_alpha2}
\alpha_{ab}=\sum_{{\bf{q}}, \lambda} C_{{\bf{q}}, \lambda} \sum_{de} S_{abde} \gamma_{{\bf{q}}, \lambda}^{de}.
\end{equation}
In a completely harmonic lattice, the frequencies would be independent of the strain and the $\gamma_{{\bf{q}}, \lambda}^{de}$ would
be zero, leading to a zero thermal expansion. Eq. \ref{def_alpha2} generalizes the expression
\begin{equation}\label{def_alpha_cubic}
\alpha=\frac{1}{3B}\sum_{{\bf{q}}, \lambda} C_{{\bf{q}}, \lambda} \gamma_{{\bf{q}}, \lambda},
\end{equation}
giving the thermal expansion coefficient $\alpha$ for a cubic solid. Here $C_{{\bf{q}}, \lambda}$ is defined in Eq. \ref{def_specific_heat} and
$\gamma_{{\bf{q}}, \lambda}$ is the mode Gr\"uneisen parameter defined as 
\begin{equation}
\gamma_{{\bf{q}}, \lambda}=-\partial \ln[\omega_\lambda({\bf{q}})]/\partial \ln V. 
\end{equation}
Usually, the phonon frequencies $\omega_\lambda({\bf{q}})$ decrease as the volume $V$ increases giving rise
to positive Gr\"uneisen parameters and thus to a positive thermal expansion coefficient $\alpha$ at any temperature as
can be inferred from Eq. \ref{def_alpha_cubic}.
Inserting the non-zero matrix elements of ${\bf{S}}$ allowed by symmetry (see Eq. \ref{compliance_tensor}) in Eq. \ref{def_alpha2} 
and taking into account the fact that $\gamma_{{\bf{q}}, \lambda}^{de}=0$ when $d\neq e$ leads to the following expressions
\begin{equation}\label{def_alpha_perp}
\alpha_\perp \equiv \alpha_{11} \equiv \alpha_{22}=\sum_{{\bf{q}}, \lambda} C_{{\bf{q}}, \lambda}
\left[\left(S_{11}+S_{12}\right) \gamma_{{\bf{q}}, \lambda}^\perp  + 
S_{13} \gamma_{{\bf{q}}, \lambda}^\parallel  \right] 
\end{equation}
and 
\begin{equation}\label{def_alpha_par}
\alpha_\parallel \equiv \alpha_{33} =\sum_{{\bf{q}}, \lambda} C_{{\bf{q}}, \lambda}
\left[ 2 S_{13} \gamma_{{\bf{q}}, \lambda}^\perp +S_{33}\gamma_{{\bf{q}}, \lambda}^\parallel  \right] 
\end{equation}
where $\alpha_\perp$ and $\alpha_\parallel$ are the thermal expansion coefficient respectively within the
basal plane and along the ternary axis. Here,
\begin{equation}\label{def_gamma_perp}
\gamma_{{\bf{q}}, \lambda}^\perp \equiv \frac{1}{2}\left(\gamma_{{\bf{q}}, \lambda}^{11}+\gamma_{{\bf{q}}, \lambda}^{22}\right) =
-\frac{a_{\perp,0}}{2 \omega^{0}_\lambda({\bf{q}})}\frac{\partial  \omega_\lambda({\bf{q}}) }{\partial a_\perp}
\end{equation}
and
\begin{equation}\label{def_gamma_par}
\gamma_{{\bf{q}}, \lambda}^\parallel \equiv \gamma_{{\bf{q}}, \lambda}^{33} =
-\frac{a_{\parallel,0}}{\omega^{0}_\lambda({\bf{q}})}\frac{\partial  \omega_\lambda({\bf{q}}) }{\partial a_\parallel}
\end{equation}
where $a_{\perp,0}$ and $a_{\parallel,0}$ are the LDA lattice parameters reported in Table \ref{lattice_tab} 
and $\omega^{0}_\lambda({\bf{q}})$ are the phonon frequencies calculated for these lattice parameters. We do not compare
our computed $\omega^{0}_\lambda({\bf{q}})$ with available experimental data because the agreement between theory and
experiments has already been highlighted in a thorough study  based on calculations performed using 
the ABINIT code\cite{diaz_2007}. A finite difference scheme based on a relative variation of $\pm 0.2$ \% of
$a_\perp$ and $a_\parallel$ around $a_{\perp,0}$ and $a_{\parallel,0}$ is used to compute the partial derivatives of
the phonon frequencies $\omega_\lambda({\bf{q}})$ with respect to $a_\perp$ and $a_\parallel$. Hence, the mode Gr\"uneisen parameters 
respectively defined in Eq. \ref{def_gamma_perp} and Eq. \ref{def_gamma_par} are computed for a $32\times 32\times 32$ qpoints grid and
for all polarizations $\lambda$.

\begin{figure}[h!]
\vskip1.5truecm
\includegraphics[width=8.5cm]{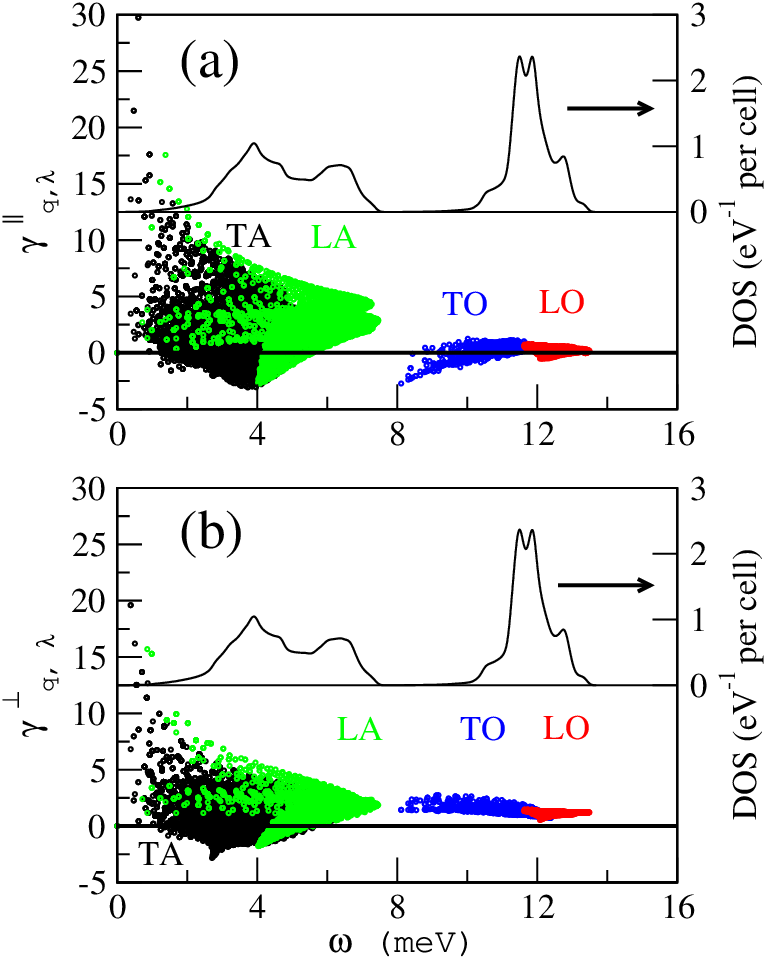}
\caption{\label{gruneisen_vs_energy}
Calculated Gr\"uneisen parameters $\gamma^\parallel_{{\bf{q}}, \lambda}$ and 
$\gamma^\perp_{{\bf{q}}, \lambda}$ as a function of phonon energy $\omega$ in meV (Left vertical scale) together with the phonon
density of states (Right vertical scale).}
\end{figure}
The Figure. \ref{gruneisen_vs_energy} shows that the mode Gr\"uneisen parameters $\gamma^\parallel_{{\bf{q}}, \lambda}$ 
and $\gamma^\perp_{{\bf{q}}, \lambda}$ are rather scattered for the acoustic modes with quite large positive values but also
negative values. About 40 \% and 18 \% of the mode Gr\"uneisen parameters are negative for the first and second transverse 
acoustic branch (TA) while less than 5 \% of the mode Gr\"uneisen parameters are negative for the longitunal acoustic branch (LA).

By comparing Figs \ref{gruneisen_vs_energy}(a) and \ref{gruneisen_vs_energy}(b), we note that the 
mode Gr\"uneisen parameters $\gamma^\parallel_{{\bf{q}}, \lambda}$ are slightly larger than the 
mode Gr\"uneisen parameters  $\gamma^\perp_{{\bf{q}}, \lambda}$ for acoustic modes with energy ranging from 0 to
7.5 meV (TA+LA). On the contrary, the mode Gr\"uneisen parameters $\gamma^\parallel_{{\bf{q}}, \lambda}$ are almost 3 times 
smaller in average than the mode Gr\"uneisen parameters $\gamma^\perp_{{\bf{q}}, \lambda}$ for optical modes with energy
ranging from 8 to 13.5 meV (TO+LO). In other words, the optical phonon frequencies are more sensitive to a variation of 
$a_\perp$ than to a variation of $a_\parallel$ while the opposite is true for the acoustic phonon frequencies.

\begin{figure}[h!]
\vskip1.5truecm
\includegraphics[width=12.0cm]{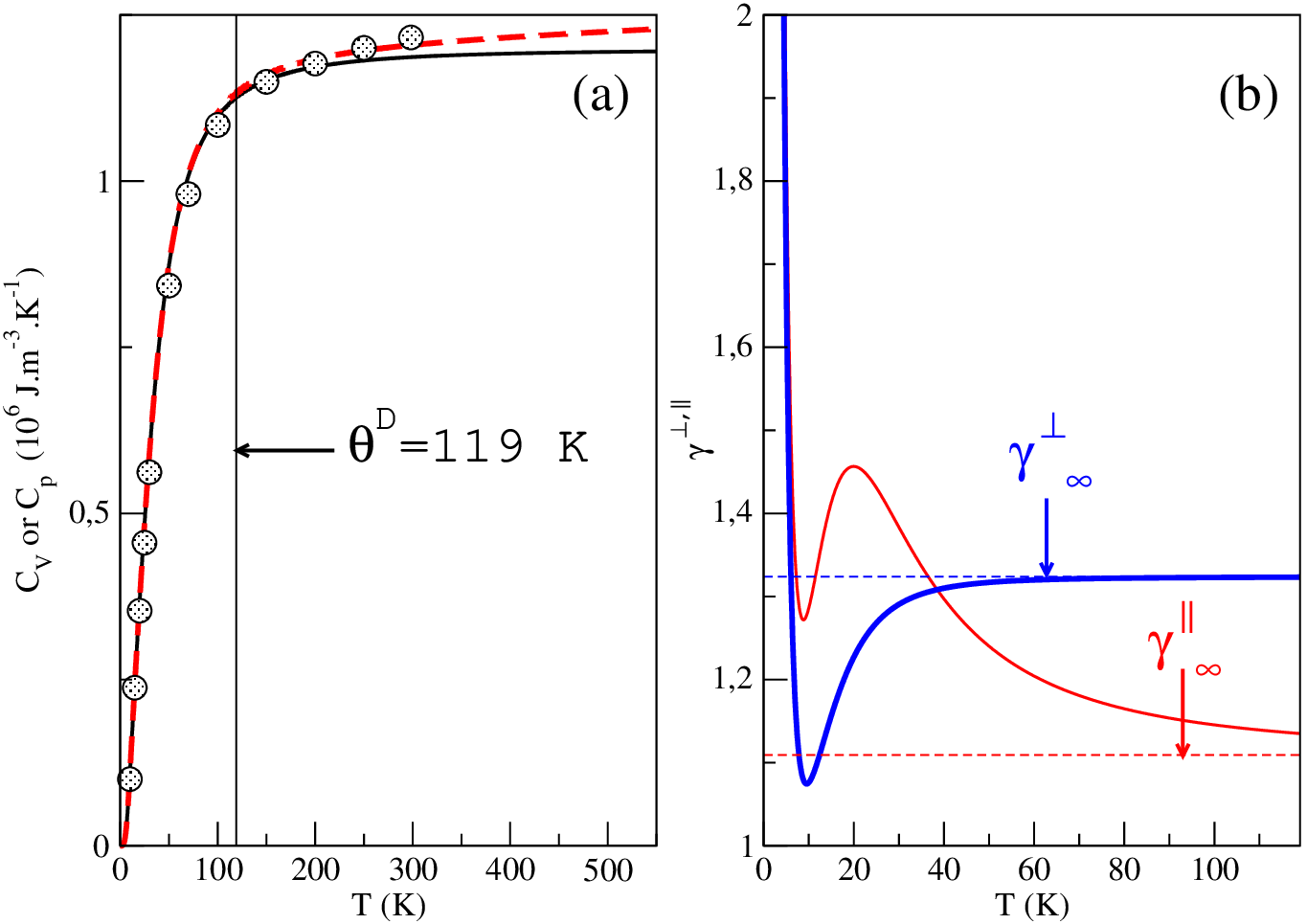}
\caption{\label{gamma_fig}
(a) Calculated lattice specific heat per unit volume at constant volume $C_V$ (solide curve) or at constant pressure $C_p$ (dashed curve) compared to experimental data (open circles) from Ref.\cite{handbook} for temperatures $T$ up to the melting temperature of 545 K. (b) Gr\"uneisen functions 
$\gamma^{\perp, \parallel}$ as a function of temperature $T$ up to the Debye temperature $\theta_D$ of 119 K\cite{handbook}.}
\end{figure}

We can also introduce macroscopic Gr\"uneisen functions
\begin{equation}\label{gamma_temp}
\gamma^{\perp, \parallel}=\left(\sum_{{\bf{q}}, \lambda} \gamma_{{\bf{q}}, \lambda}^{\perp, \parallel} C_{{\bf{q}}, \lambda}\right)
/C_V
\end{equation}
where $C_{{\bf{q}}, \lambda}$ and $C_V$ are respectively defined in Eq. \ref{def_specific_heat}
and \ref{def_specific_heat_2}. The calculated lattice specific heat at constant volume $C_V$ displayed in
Fig. \ref{gamma_fig}(a) is in good agreement with the experimental lattice specific heat at constant pressure
up to the Debye temperature $\theta_D=119$ K\cite{handbook}. As shown in Fig. \ref{gamma_fig}(b), the behaviour of the temperature dependent 
Gr\"uneisen functions $\gamma^{\perp}$ and $\gamma^{\parallel}$ is quite complex with a crossover around 40 K. 
However, $\gamma^{\perp}$ and $\gamma^{\parallel}$ saturate towards $\gamma^{\perp}_\infty=1.32$ and
$\gamma^{\parallel}_\infty=1.11$ when $T\gg\theta_D$ in accordance with the fact that
\begin{equation}
\lim_{T\to\infty} \gamma^{\perp, \parallel}(T)=\frac{1}{6N} \sum_{{\bf{q}}, \lambda} \gamma_{{\bf{q}}, \lambda}^{\perp, \parallel},
\end{equation}
since $C_{{\bf{q}}, \lambda}(T) \to k_B/V_0 N$ and $C_V(T) \to 6 k_B/V_0$ when $T\to\infty$. Interestingly, the high temperature
limiting values of $\gamma^{\perp}_\infty$ and $\gamma^{\parallel}_\infty$ extracted from experimental results\cite{bunton_1969}
are estimated to be 1.32 and 1.10 and are in excellent agreement with our calculated values.

\begin{figure}[h!]
\vskip1.5truecm
\includegraphics[width=10.0cm]{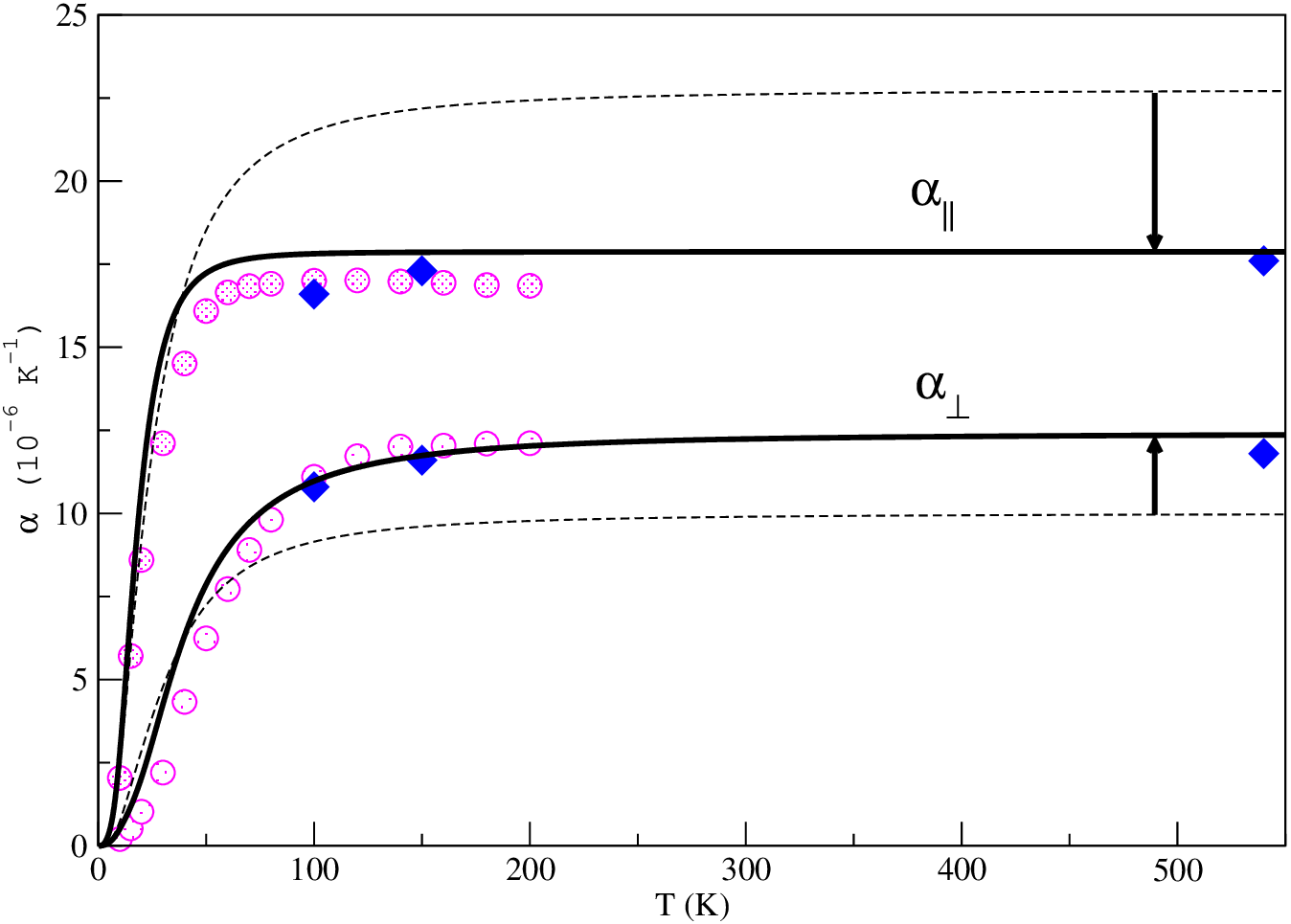}
\caption{\label{dilatation_fig}
Coefficient of linear thermal expansion of bismuth within the basal plane ($\alpha_\perp$) 
and along the ternary axis ($\alpha_\parallel$) as a function of temperature T 
up to the melting temperature. Experimental data are denoted by circles\cite{bunton_1969} and losanges\cite{cave_1960}. 
The full curves are calculated using Eq. \ref{def_alpha_perp_new} and \ref{def_alpha_par_new} while the dashed curves are
calculated by neglecting the anisotropy in the Gr\"uneisen functions, which is tantamount to put $S_{13}=0$ 
in Eq. \ref{def_alpha_perp_new} and \ref{def_alpha_par_new}.}
\end{figure}

Following the approach of Munn\cite{Munn_1972}, the principal coefficients of thermal expansion defined 
in Eq. \ref{def_alpha_perp} and \ref{def_alpha_par} can also be expressed as
\begin{equation}\label{def_alpha_perp_new}
\alpha_\perp = C_V \left[\chi_\perp \gamma^{\perp} + S_{13} \left(\gamma^{\parallel}-\gamma^{\perp}\right)    \right] 
\end{equation}
\begin{equation}\label{def_alpha_par_new}
\alpha_\parallel = C_V \left[\chi_\parallel \gamma^{\parallel} - 2 S_{13} \left(\gamma^{\parallel}-\gamma^{\perp}\right)    \right] 
\end{equation}
where the directional Gr\"uneisen functions $\gamma^{\perp, \parallel}$ are defined in Eq. \ref{gamma_temp} and
the compressibilities $\chi_\parallel$ and $\chi_\perp$ are respectively defined in Eq.  \ref{comp_par} and \ref{comp_perp}. It is
a good first approximation to treat both the compressibilities and the compliance matrix element $S_{13}$ as constant, 
and regard the temperature dependence of the coefficients of thermal expansion
as due solely to that of the heat capacity $C_V$ and the Gr\"uneisen functions $\gamma^{\perp, \parallel}$.  
Such expressions for the coefficients of thermal expansion
allow to disentangle the role of the anisotropy in either the Gr\"uneisen functions or the elastic constants. Fig. \ref{gamma_fig}(b)
shows that $\gamma^{\parallel}=\gamma^{\perp}$ for T=40 K. Thus, the anisotropy in the thermal expansion coefficients measured
by $\alpha_\parallel/\alpha_\perp$ is given by $\chi_\parallel/\chi_\perp=2.71$ (see Table \ref{compressibility_tab}) and
is only due to the anisotropy in the elastic properties. When moving away from the crossover temperature (T=40 K), the term
proportionnal to $\gamma^{\parallel}-\gamma^{\perp}$ starts to play a role as the cross-compliance $S_{13}=-C_{13}/M$ ($\sim$ -0.94 MBar$^{-1}$) 
has the same order of magnitude as $\chi_\parallel$ and $\chi_\perp$ (see Table \ref{compressibility_tab}). 
For $\alpha_\parallel$, the correction arising from the anisotropy in the Gr\"uneisen functions is given by 
$-2 C_V S_{13} \left(\gamma^{\parallel}-\gamma^{\perp}\right)$ and remains very small at low temperature since $C_V \to 0$ when
$T \to 0$. Thus, the correction is small and positive for $T<40$ K since $S_{13}<0$ and $\gamma^{\parallel}-\gamma^{\perp}>0$ in
this low temperature regime. However, the correction becomes non negligible at higher temperatures and negative as the sign
of $\gamma^{\parallel}-\gamma^{\perp}$ changes when $T>40$ K. The high temperature limit of this correction is given
by $-12 k_B S_{13} (\gamma^{\parallel}_\infty-\gamma^{\perp}_\infty)/V_0$ and is sketched as a vertical downward arrow 
in Fig. \ref{dilatation_fig}. For $\alpha_\perp$, the same type of conclusion holds but the sign and the magnitude of the
correction is changed since it is given by $C_V S_{13} \left(\gamma^{\parallel}-\gamma^{\perp}\right)$. 
The high temperature limit of this correction is given by $6 k_B S_{13} (\gamma^{\parallel}_\infty-\gamma^{\perp}_\infty)/V_0$ 
and is depicted as a vertical upward arrow in Fig. \ref{dilatation_fig} since this quantity is positive.
The thermal expansion coefficients parallel ($\alpha_\parallel$) and  perpendicular ($\alpha_\perp$) to the ternary axis calculated
neglecting (dashed curves) and including (full curves) the anisotropy in the Gr\"uneisen functions are displayed in Fig. \ref{dilatation_fig}.
At low temperature ($T<40$ K), the anisotropy in the Gr\"uneisen functions plays a minor role since the full curves almost coincide with the
dashed curves. However, it starts to contribute at higher temperature since it reduces $\alpha_\parallel$ and increases $\alpha_\perp$ significantly, thereby reducing the anisotropy in the thermal expansion coefficients given by $\alpha_\parallel/\alpha_\perp$ and
bringing our calculated thermal expansion coefficients in closer agreement with the measurements made using 
an optical lever dilatometer\cite{bunton_1969, cave_1960}. Thus, the quasi-harmonic approximation based on quantities calculated at 0 K provides
quite accurate results compared to experimental data\cite{bunton_1969, cave_1960}, even near the melting temperature of bismuth where renormalization
effects due to the temperature dependence of elastic constants might be important\cite{debernardi_2001} and where the applicability of the quasiharmonic
approximation is also questionable because large anharmonic effects are expected. It's worth coming back to the calculated lattice specific heat 
at constant volume $C_V$ shown in Fig. \ref{gamma_fig}(a). 
Above $\theta_D$, the experimental lattice specific heat at constant pressure deviates from $C_V$ and should in principle be compared to $C_p$. A very
simple reasoning based on thermodynamics\cite{callen} shows that
\begin{equation}
C_p-C_V=\alpha^2 B T,
\end{equation}
where $\alpha\equiv\frac{1}{V_0} \left(\frac{\partial V}{\partial T}\right)_P=\alpha_{\parallel}+2\alpha_{\perp}$ is the volumetric expansion coefficient 
and $B$ is the temperature dependent bulk modulus, which can be approximated by it's 0 K value. The calculated $C_p$ displayed in Fig. \ref{gamma_fig}(a)
coincides with the calculated $C_V$ below the Debye temperature $\theta_D$ because $\alpha(T)\to 0$ when $T\ll \theta_D$, making the correction 
$\alpha^2 B T$ very small in this temperature range. When $T>\theta_D$, $\alpha(T)$ becomes constant and the correction increases linearly with $T$. The
agreement between theory and experiment is improved since all the experimental data (open circles) collapse on the calculated $C_p$ (dashed curve) up
to room temperature.   

\section{conclusion}
We performed first-principles calculations in order to understand the anisotropic thermal expansion of bismuth. First, we computed
the elastic constants of bismuth at 0 K using a finite strain approach. All the elastic constants, with the exception of the $C_{14}$ and $C_{44}$ elastic constants, are found to be in good agreement with experimental results\cite{lichnowski_1976} when the SOI is included.
We also calculated the hydrostatic compressibilities along the ternary axis ($\chi_{\parallel}$) and perpendicular to it ($\chi_{\perp}$)
and found that the anisotropy in the directional compressibilities is large since $\chi_{\parallel}/\chi_{\perp}\sim 2.7$.
Then, we computed the thermal
expansion coefficients parallel ($\alpha_{\parallel}$) and perpendicular ($\alpha_{\perp}$) to the ternary axis using the quasiharmonic 
approximation. These quantities are found to be in close agreement with experiments\cite{bunton_1969, cave_1960}. Another outcome of our
calculations is that the anisotropy in the thermal  expansion coefficients is essentially governed by the anisotropy in the mechanical properties
below the Debye temperature $\theta_D$ while both the anisotropy in the directional compressibilities and in the directional Gr\"uneisen functions 
play a role at higher temperatures. Finally, this work is a first step towards a first-principles description of the thermal/non thermal expansion
in laser-excited bismuth\cite{laulhe_2013}, where the electron system is not equilibrated with the phonon system\cite{giret_2011, arnaud_2013}. 

\begin{acknowledgments}
Calculations were performed using HPC resources from GENCI-CINES (project 095096).
\end{acknowledgments}

\end{document}